\documentclass[aps, prl, 10pt, twocolumn,superscriptaddress]{revtex4-1}

\usepackage{epsfig}
\usepackage{graphicx}
\usepackage{float}
\usepackage{dcolumn}
\usepackage{xcolor}
\usepackage{bm}
\usepackage{amsmath,bbm}

\begin{document}

\title{Quest for topological superconductivity at superconductor-semiconductor interfaces}

\author{S.M. Frolov}
\affiliation{Department of Physics and Astronomy, University of Pittsburgh, Pittsburgh, PA 15260, USA} 
\author{M.J. Manfra}
\affiliation{Department  of  Physics and Astronomy,  and Microsoft Quantum Purdue, Purdue  University,  West  Lafayette,  Indiana  47907,  USA} 
\author{J.D. Sau}
\affiliation{Department of Physics, University of Maryland, College Park, Maryland 20742-4111, USA} 

\date{\today}

\begin{abstract}
We analyze the evidence of Majorana zero modes in nanowires that came from tunneling spectroscopy and other experiments, and scout the path to topologically protected states that are of interest for quantum computing. We illustrate the importance of the superconductor-semiconductor interface quality and sketch out where further progress in materials science of these interfaces can take us. Finally, we discuss the prospects of observing more exotic non-Abelian anyons based on the same materials platform, and how to make connections to high energy physics.
\end{abstract}

\maketitle

Topological superconductivity is distinct from other kinds of superconductivity in subtle yet profound ways. What makes it a crown jewel among topological phases is its capacity to harbor non-Abelian (i.e. neither fermionic nor bosonic) states  \cite{Read2000,Ivanov2001}. Establishing new classes of particles is of clear fundamental interest, but further impact may come from applications in fault-tolerant quantum computing \cite{KitaevAP03, NayakRMP2008, sternscience2013, sarmanjp2015}. Topological superconductivity can emerge intrinsically in the bulk of a material \cite{volovik1999fermion}, or it can be induced at an interface between two materials \cite{Fu2008PRL}. Among the incredible variety of proposed hosts, here we focus on interfaces of two of the most common materials - a superconducting metal and a semiconductor. Even though neither of the constituents is topological by itself, topological superconductivity is predicted when ingredients such as particle-hole symmetry and spin-orbit interaction are borrowed from either side of the interface \cite{SauPRL2010, AliceaPRB2010, LutchynPRL2010, OregPRL2010}. The power of this approach is that it offers a relatively straightforward pathway to generating, detecting and manipulating Majorana zero modes (MZMs) - the most basic of non-Abelian anyons \cite{KitaevPU2001,Sengupta2001,wilczeknphys2009}.

\begin{figure}[t!]
\centering
  \includegraphics[width=0.45\textwidth]{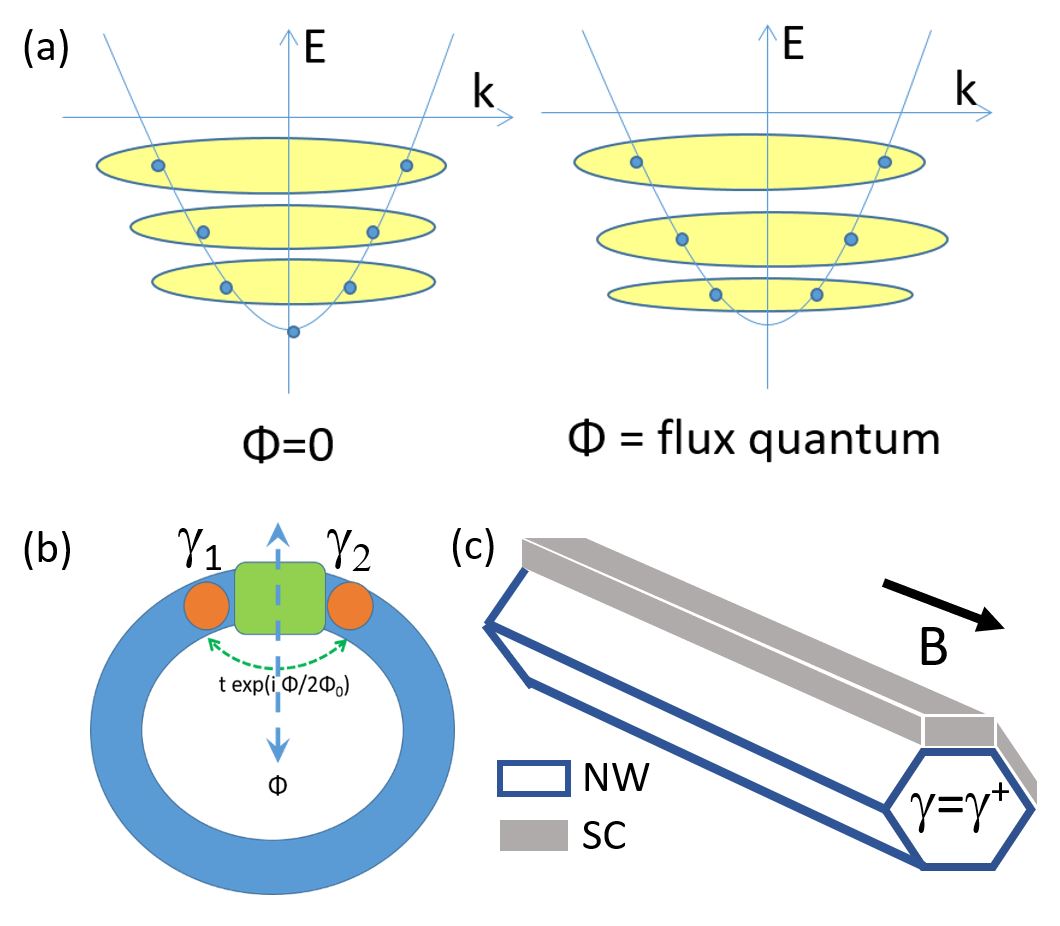}
\caption{ The concept of topological superconductivity using a spinless superconductor primer.
(a) Fermion parity anomaly in a ring of a spinless superconductor. Under zero external flux threading the ring, the ground state of a spinlesss superconductor features a single unpaired fermion at k=0. When a flux of a single flux quantum is applied, the ground state is fully paired with even total parity. (b) MZM (orange circles) are nucleated in a spinless superconductor ring when a tunnel barrier (green) is introduced, with a flux-dependent tunneling amplitude t($\Phi$). (c) the ingredients of an effective spinless superconductor considered here are the strong spin-orbit semiconductor nanowire (NW) coupled to a conventional superconductor (SC) in an external magnetic field B. MZMs $\gamma$ are expected at the ends of the nanowire.
 \label{fig1}}
\end{figure}

In fact, a traditional way of introducing topological superconductivity is through Majorana zero modes which are edge, end or defect states \cite{leijnsesst2012, AliceaPRP2012, BeenakkerARCMP2013, stanescujpcm2013, aguadornc2017, lutchynnrm2018}. Here, we draw attention to a unique property of the topological phase as a whole: the fermion parity anomaly which is a hallmark of time-reversal symmetry-breaking topological superconductors \cite{Read2000,KitaevPU2001}. We shall then motivate how this anomaly directly mandates MZMs at system boundaries. Let's start by considering a system without boundaries: a ring of a strictly one-dimensional and spinless superconductor. Fermions, of course, possess half-integer spin, but theoretically we are free to assume spinless particles whose creation operators still anticommute \cite{KitaevPU2001}. The wavefunctions are periodic around the ring, and momenta take discrete values including one at k=0 when flux through the ring is zero (Fig. 1a). Because these fermions have no spin, the state at k=0 has no same-energy partner to form a Cooper pair with. We have thus realized a superconducting state with a single unpaired electron. This is already unusual for conventional (non-topological) superconductors which are fully paired in the ground state. The fermion parity anomaly is exposed when we apply a quantum of magnetic flux through the ring. The boundary conditions change and the k=0 state disappears: all electrons now have a Cooper pair partner (Fig. 1a). The fermion parity of the ground state has changed from odd to even: this is an anomaly. It has not been directly observed but may in principle be accessed through Coulomb blockade oscillations of conductance \cite{liuprl2019}.

The genesis of Majorana modes takes place when the spinless superconducting ring is cut open, and thereby transformed into a one-dimensional wire. Let us do this gradually by inserting a barrier into the ring (Fig. 1b). As long as the barrier allows tunneling, the fermion parity anomaly must persist, and the ground state parity should change upon insertion of flux. But the flux through the loop now controls the phase of the tunneling amplitude: changing flux by a single quantum flips the tunneling amplitude from $+t$ to $-t$. Transitioning from an even- to odd-parity ground state then requires there to be states with energies of order $t$~\cite{KitaevPU2001}. In the limit of infinitely small $t$ these are zero energy states bound to the ends as a consequence of topological fermion parity anomaly, they are the MZMs.

While spinless fermions do not exist, one strategy to obtain an effectively spinless superconductor is to spin-polarize all electrons. But this destroys conventional spin-singlet superconductivity making the most common superconductors not suitable. A workaround is to induce superconductivity in a spin-polarized semiconductor by proximity to a spin-singlet superconductor (Fig. 1c) \cite{SauPRL2010,SauPRB2010, AliceaPRB2010, LutchynPRL2010, OregPRL2010}. External magnetic field required to polarize semiconductors such as InAs or InSb, with their large effective g-factors, is low enough to preserve superconductivity in thin films of Al or Nb. And spin-orbit interaction within InAs or InSb is large enough to break the perfect spin-polarization serving as a bridge between singlet and triplet Cooper pairing.

In order to localize MZMs it is necessary to implement a one-dimensional system, such as a nanowire (Fig. 1c). Initially semiconductor nanowires grown via vapor-liquid-solid (VLS) phase epitaxy   \cite{law2004semiconductor} coupled to s-wave superconductors ex-situ (in a separate fabrication process) were used. But new approaches to growth such as integrating epitaxial superconducting layers in-situ \cite{krogstrup2015, Shabani2016,Gazibegovicnature2017}, and developing planar superconductor/2DEG hybrids as well as selective area growth (SAG) of superconductor-semiconductor wire networks  \cite{sugaya1992selective, nishinaga2000selective} may advance progress towards unambiguous demonstration of MZMs, and facilitate experiments requiring more complex geometries.

Superconducting proximity effects in low dimensional semiconductors have been studied for a two decades prior to the first 2012 wave of Majorana experiments \cite{defranceschinatnano2010}, but primarily close to zero applied magnetic field, thus not in the spin-polarized regime. Following the 2010 Majorana predictions \cite{LutchynPRL2010, OregPRL2010}, finite field experiments were performed on semiconductor nanowire devices. They largely focused on identifying zero voltage bias conductance peaks (ZBCP’s) \cite{MourikScience2012, DasNatphys2012, DengNanolett2012, FinckPRL2013, ChurchillPRB2013}. It was quickly realized that the observed peaks did not clearly correspond to any previously known phenomenon (Fig. 2a). Their most striking feature was ZBCP pinning to zero bias voltage upon significant changes in magnetic field. Even though they appeared only at finite magnetic field, the resonances apparently lacked Zeeman energy, i.e. behaved as spinless, zero-energy states - just as Majorana modes should.

\begin{figure}[t!]
\centering
  \includegraphics[width=0.48\textwidth]{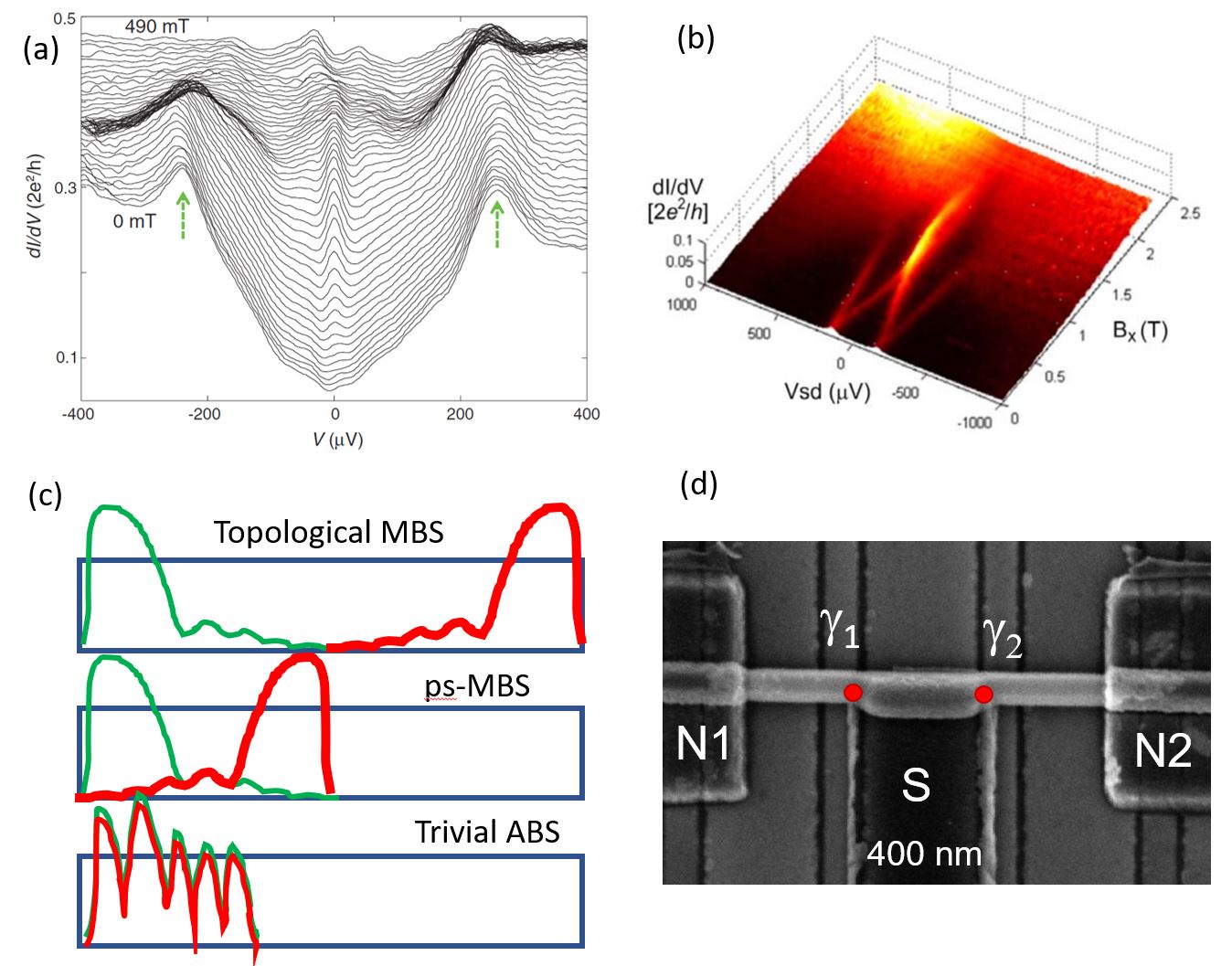}
\caption{Zero-bias conductance peaks can appear due to MZM or due to trivial ABS. (a) ZBCP emerging at finite magnetic field as reported by one of the 2012 nanowire experiments \cite{MourikScience2012}. (b) Another ZBCP appearing at a confluence of two resonances at a finite field. These data appear in the supplementary materials of Ref. \cite{LeeNatnano2014} where a quantum dot device not capable of sustaining MZM was studied. (c) Subgap wavefunctions in a semiconductor nanowire coupled to a superconductor decomposed in left Majorana (green) and right Majorana (red) basis. Three situations are presented: well-separated topological MZM (top), partially separated Andreev bound states (middle), completely trivial ABS (bottom) where green and red wavefunctions fully overlap. (d) A three-terminal nanowire device designed to probe the left and right end of a superconductor-semiconductor segment S via normal probes N1 and N2 (Yu, Bakkers, Frolov, unpublished).
 \label{fig2}}
\end{figure}

The field took a surprise swing when resonances that similarly did do not shift from zero energy in magnetic field were identified due to Andreev bound states (ABS) in quantum dot devices (Fig. 2b)~\cite{LeeNatnano2014}. The similarities between ABS and MZM run deep: both phenomena correspond to low energy (subgap) bound states near superconductor boundaries.  A long topologically superconducting wire is expected to have two strictly zero energy MZMs at the ends. In a short wire, the two MZM partially overlap, acquire non-zero energy and therefore become ABS. In turn, regardless of whether the wire is topological or not, any ABS can be represented in the ‘Majorana basis’ or in other words split into two Majorana wavefunctions (‘left’ and ‘right’) (Fig. 2c). The two wavefunctions may fully overlap which is the case for trivial ABS~\cite{Liu2017a}, or they may strongly overlap which is known as partially-separated Andreev or quasi-Majorana states \cite{Moore2018a,Stanescu2018b, vuik2018}. A smooth nature of the barrier potential plays a crucial role in keeping such quasi-Majorana states near zero energy~\cite{KellsPRB2012,Liu2018}. What is interesting is that quasi-Majoranas do not need to accompany a bulk topological phase, but may arise generically under similar conditions of strong spin-orbit coupling and large magnetic field~\cite{vuik2018,pan2019generic}. An alternative source for non-topological ZBCPs is referred to as 'class-D' peak~\cite{BrouwerBeenakker,AltlandZirnbauer,PikulinNJP2012,Bagrets2012}. It arises due to the level repulsion from other nearby states pushing some resonances to low bias. While class-D peaks generically exhibit less zero-bias pinning than MZM, it is possible to find instances of considerable pinning through data selection~\cite{Woods2019, pan2019generic}. 

In light of these realizations, more nuanced attributes predicted for Majorana ZBCPs were considered in experiments on improved nanowire devices which aimed to eliminate or separate out trivial ABS. Signatures such as $2e^2/h$ ‘quantized’ conductance, topological phase diagram in Zeeman energy and chemical potential, near-zero bias oscillations of conductance resonances, degree of zero-bias pinning for different length segments, closing of the apparent superconducting gap, etc. were considered \cite{Albrecht2016,Deng2016,Zhang2017,Chen2017,Nichele2017,Zhang2018, Kjaergaard2016, grivnin2019concomitant}. While each observation was consistent with MZM, many were also reproduced without assuming MZM \cite{Moor2018, ChenPrl19, pan2019generic}. Similar fate was in store for the efforts to demonstrate the fractional Josephson effect \cite{RokhinsonNPhys2012, laroche2019observation} which at first was widely believed to be unique to MZM. It has been realized that this effect may not appear in a topological system due to dynamics effects such as quasiparticle relaxation and Landau-Zener transitions \cite{LutchynPRL2010, houzetprl2013}, the latter also responsible for apparent fractional Josephson effect even in non-topological systems \cite{BillangeonPRL07,SauBerg,SetiawanShapiro}. 

While collectively ZBCPs were explored extensively, each individual experiment demonstrates only one or two of the predicted signatures of MZM, while not at the same time finding all expected features. Given that the similarities between ABS and MZM are so striking, the prospects of coming up with a clear single-figure 'smoking gun' evidence of MZM is unlikely. Instead, a comprehensive set of internally consistent measurements performed on the same sample will be needed to unambiguously establish the existence of Majorana modes. One advantage of the super-semi system is in the large number of control parameters that allow for elaborate testing of the MZM hypothesis: through gate voltages, magnetic field magnitude and orientation, device geometry, materials and interface tailoring. 

One property which may, if clearly demonstrated, definitively distinguish ABS from MZM is the non-local nature of Majorana wavefunctions, which is the correlation between left and right MZM. Majorana nonlocality has been indirectly explored in two-terminal devices~\cite{DengPRB18}, and first experiments began to appear in three-terminal  devices (Fig. 2d) \cite{grivnin2019concomitant, anselmetti2019end}, where near-zero energy states on the two ends of a nanowire can be probed independently. Alternatively, such three-terminal devices can be used to probe signatures of the topological quantum phase transition such as quantized heat conductance~\cite{Akhmerovquantized} and 
non-local rectifying electrical conductance~\cite{AkhmerovRectifier}. Proposed Majorana teleportation and noise correlation measurements as well as Majorana mode interferometry are some of the future experiment with large potential \cite{FuPRL2010, SauSwingle,MichaeliFu}.

Topological superconductivity at superconductor-semiconductor interfaces is strongly dependent on the materials and interface properties. In contrast with conventional superconductivity, topological superconductivity is highly sensitive to any disorder, including scattering on non-magnetic impurities \cite{Sau2012}. Intrinsic material parameters such as band offsets at the interfaces, which are not well-known at present, may influence whether the system is close to or far from the theoretical single one-dimensional channel limit. The effective g-factor and spin-orbit coupling at the super/semi interface may also be renormalized, impacting the formation and stability of a topological phase. Lattice-mismatch and changes of crystal symmetry between the superconductor and semiconductor are important. While these concerns are specific to super-semi systems, we expect detailed materials considerations to be key for all attempts to implement topological superconductivity.

One of the recent highlights with large potential for the future is the development of two dimensional superconductor-semiconductor heterostructures. Clean interfaces are created between an s-wave superconductor (typically aluminum) and a two-dimensional electron gas with strong spin-orbit coupling within the ultra-pure molecular beam epitaxy (MBE) environment (Figs. 3a-3c)\cite{Shabani2016, Kjaergaard2016}. The superconductor and the semiconductor can be separated by a thin  tunneling barrier  to control the induced superconductivity. 2D growth of InAs or InSb generally involves heteroepitaxy on lattice-mismatched substrates, as those are the only available insulating options. Extended defect generation is inevitable and must be effectively managed if device quality is to be maintained. First devices made out of these heterostructures featured one-dimensional channels defined either via top-down etching, or within a Josephson junction \cite{Nichele2017, ren2019topological, fornieri2019evidence}. These experiments obtained ZBCPs qualitatively similar to those observed in VLS nanowire-based devices, leaving the issue of separating MZM and ABS open for the moment. 

Sustained progress will rely on deepening our understanding of electronic properties specific to superconductor-semiconductor interfaces using a combination of theory and also spectroscopy tools beyond electron transport. It is expected that the established strengths of MBE growth of semiconductor heterostructures, including in-situ diagnostics, high chemical purity, interface control with monolayer precision may be leveraged to yield even lower defect density and interfaces maximally optimized for MZM generation.

\begin{figure}[t!]
\centering
  \includegraphics[width=0.48\textwidth]{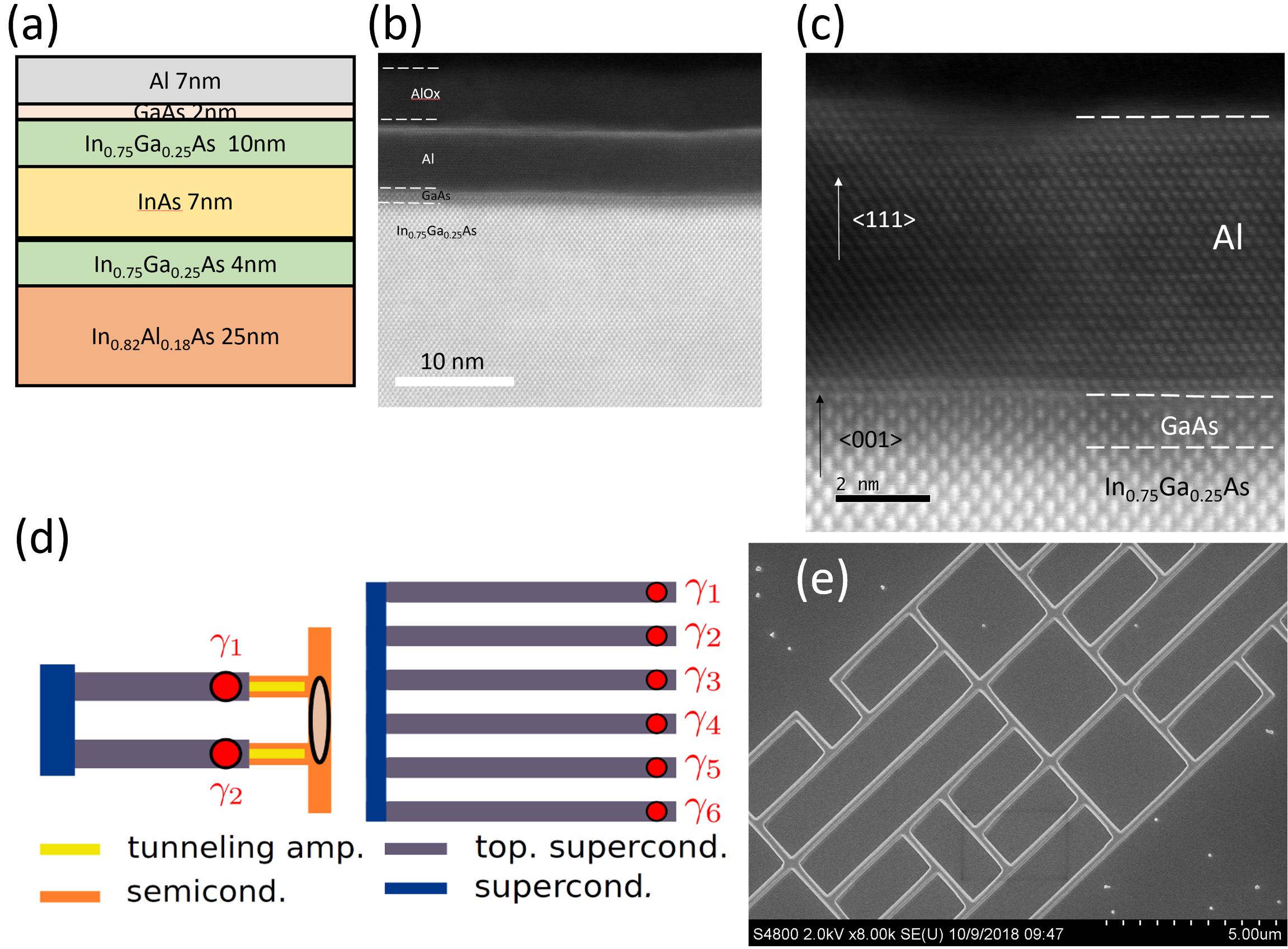}
\caption{Material considerations for superconductor/semiconductor hybrid systems.
(a) Schematic stack of planar Al/InAs 2DEG structure similar to that used in Ref. \cite{NichelePRL2017}. (b) Scanning transmission electron microscopy (STEM) image of the InAs quantum well region, including 10nm InGaAs barrier and 7nm epitaxial aluminum. (c) High resolution STEM image demonstrating epitaxial arrangement at the superconductor/semiconductor interface.  The (111) planes of aluminum are parallel to the (001) planes of the III-V semiconductor, transitioning at an abrupt interface. Images in (b)-(c) courtesy of R.E. Diaz. (d) Schematic building blocks of topological nanowire networks required for proposed topological quantum computing schemes as in Ref. \cite{Karzig2017}. (e) Scanning electron microscopy image of an InAs nanowire array grown by SAG demonstrating the capacity to create complex networks (Gronin, Manfra, unpublished).
 \label{fig3}}
\end{figure}

Undeniably, what continues to fuel interest in topological superconductivity is the predicted non-Abelian statistics of Majorana zero modes, and its potential utility for quantum information processing \cite{KitaevAP03, NayakRMP2008, sternscience2013}. The most intuitive way of understanding this property is by physically moving two MZM around each other and observing that the ground state parity has changed. Quantum information stored in charge parity is protected against small perturbations for a subset of quantum logic operations stopping short of universal quantum computing. Moving MZM is technically  challenging for Majorana modes bound to ends of wires. Fortunately, it can be emulated by a combination of auxiliary Majorana wires together with measurements of the fermion parity, i.e. the so-called measurement-only approaches \cite{AliceaNatphy2011, vanHeckNJP2012,Karzig2017, stenger2019braiding}. The variety and increasing level of detail in these proposals clearly sets the super-semi platform apart in the efforts to realize scalable topological quantum computing.  The envisaged geometries, however, are quite complex and difficult to realize with VLS nanowires or etching 2D heterostructures (Fig. 3d). Here selective-area growth (SAG) strategy, if proven to produce material of sufficient quality, may find utility as it front loads the fabrication effort prior to epitaxial growth (Fig. 3e) \cite{sugaya1992selective, nishinaga2000selective, Peter2018PRM, aseev2018selectivity, lee2019selective, friedl2018template}. The predetermined network is grown through a patterned dielectric mask deposited on a semiconductor substrate, superconductors may be then deposited in-situ on predetermined wire facets. Exploration of SAG capabilities and limitations is an active area of research. 

Coherence of a Majorana qubit depends on the magnitude of the induced superconducting gap which is a measure of topological protection. It is also important to avoid non-equilibrium effects known as quasiparticle poisoning, which lead to fluctuations in charge parity and thereby scramble the two states of a Majorana qubit \cite{rainis2012majorana}. To date, aluminum is the preferred superconductor because of its relative resistance to quasiparticle poisoning. Indeed, in Coulomb-blockaded devices aluminum is the only known superconductor to produce 2e-periodic transport, which does not alter charge parity \cite{LafargePRL93, Albrecht2016}. The superconducting gap of aluminum at zero magnetic field is 200$\mu$eV (equivalent to 2 K) which gives the upper bound on any induced topological gap. Material parameters inherent to real superconductor/semiconductor interfaces will tend to reduce the gap in the topological regime, perhaps significantly. Thus there is strong motivation to discover/explore other superconductors that may enhance the topological gap while maintaining aluminum’s desirable characteristic of 2e-periodic transport \cite{pendharkar2019parity, bjergfelt2019superconducting, carrad2019shadow}.

The same superconductor-semiconductor interfaces inspire rich topological physics that goes beyond Majorana zero modes, and may offer pathways to universal topological quantum computing~\cite{barkeshlisau} or to quantum simulation of fundamental phenomena such as supersymmetry~\cite{OregSUSY} and black holes~\cite{AliceaSYK}. A great example is offered by parafermions~\cite{ClarkeParafermions,Lindnerparafermions,Chengparafermions,AliceaParafermions}, which are generalizations of Majorana modes where a non-interacting nanowire is replaced with a fractional quantum Hall edge state (such as states at filling factors $\nu = 1/3,2/3,4/3...$)~\cite{AliceaParafermions}. Parafermions can be intuitively thought of as fractionalized Majorana modes. They obey more complex non-Abelian rules that can perform the entire set of Clifford gates even without measurement. Parafermions have not been realized, but work on this topic begins with well-characterized superconductor-semiconductor two-dimensional interfaces. In order to enter the quantum Hall regime, superconductivity must withstand large out-of-plane magnetic fields, ruling out the use of aluminum with its critical field of order 10 mT. This provides another reason to expect many other super/semi combinations to be tried in the near future. Van der Waals heterostructures featuring materials such as graphene and layered superconductors also hold great promise for the realization of parafermions, with induced superconductivity in the quantum Hall regime already demonstrated~\cite{FinkelsteinQHSC,KimQHSC}.

Topological superconductivity with the addition of Coulomb interactions has been a highly motivating topic for theorists and may lead to interesting experiments. Here we highlight an Ising topological phase~\cite{Moore1991,Kitaevhoneycomb,YaoKivelson} built from an array of interacting nanowires (Fig. 4a). This phase differs qualitatively from two-dimensional topological superconductors in having visons, the vortices that bind Majorana modes. Visons are truly localized in contrast with Abrikosov or Josephson vortices that are associated with a non-local halo of phase winding. The Ising topological phase is interesting because it supports topological degeneracy, which arises from arranging these Majorana wires in a circuit with non-trivial topology even if there are no free ends. The Ising topological phase also leads to completely topologically protected quantum computing~\cite{barkeshlisau}.

\begin{figure}[t!]
\centering
  \includegraphics[width=0.48\textwidth]{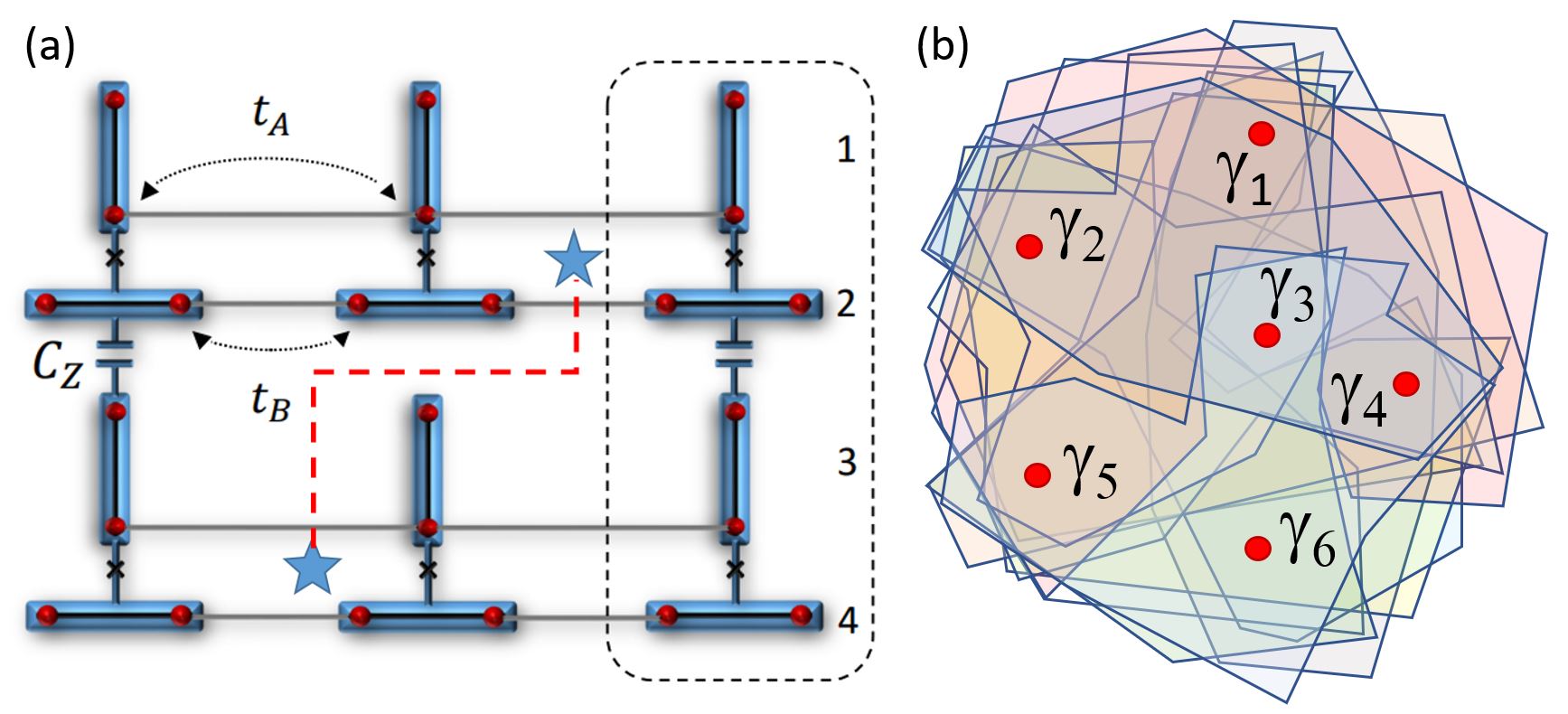}
\caption{
(a) Array of Coulomb blockaded Majorana wires (MZMs are red dots) can be used to modulate the sign of the Majorana tunneling leading to an emergent $\textbf{Z}_2$ gauge-field similar to an Ising topological phase. Sign-flip hopping around the red-dashed lines leads to a pair of local excitations (visons) shown by stars.~\cite{barkeshlisau} (b) A cartoon representation of the SYK model which requires randomized four-way interactions (polygons) to dominate in a system of many MZMs (red circles, $\gamma$'s)~\cite{AliceaSYK}. 
 \label{fig4}}
\end{figure}

What lies ahead is a unique opportunity to realize and study supersymmetry on a semiconductor chip~\cite{WessBagger}. Coulomb interactions in a one-dimensional chain lead to phase fluctuations, inviting the possibility of a topological version of a superconductor-insulator transition. It has recently been shown theoretically~\cite{OregSUSY} that in specific limits, this transition can coincide with the topological superconducting phase transition leading to a combined supersymmetric critical point~\cite{FriedanSUSY}. A symmetry that emerges at the critical point should connect the bosonic phase mode and the fermionic quasiparticle mode. 

Another exotic phase deserving of attention is the so-called Sachdev-Ye-Kitaev (SYK) model~\cite{SY,KitaevSYK} based on four-Majorana interactions. This phase may be realized when a large number of Majorana modes is forced to remain at zero energy by a special chiral symmetry (Fig. 4b). Coupling a bundle of wires to a disordered quantum dot can induce random Coulomb interactions between the MZMs that ultimately realizes the SYK model~\cite{AliceaSYK}. The SYK model is remarkable from the theoretical point of view because it is one of the few strongly interacting solvable models that thermalizes. The model is particularly interesting as it scrambles quantum information at the maximal rate that is allowed\cite{Maldacenabound}. This makes the behavior of this system relate to conjectured quantum mechanical properties of  black holes making it in some sense possible to realize a black hole in a table-top experiment. 

Clearly the effort to realize Majorana modes, quite apart from the potential for quantum computation, will likely have broader implications for deeper concepts in quantum many-body physics. The path forward lies through improved understanding of materials science of super/semi interfaces, advanced quantum engineering and methodical partnership between experiment and theory.

Acknowledgements. The authors thank S. Das Sarma for comments. S.M.F. is supported by NSF DMR-1743972, NSF DMR-1906325, NSF PIRE-1743717, ONR and ARO. M.J.M is supported by Microsoft Quantum. 

\bibliographystyle{apsrev4-1}
\bibliography{Majorana_NPhys.bib}

\end{document}